\documentclass[12pt]{iopart}

\usepackage[hidelinks]{hyperref}
\usepackage{iopams}
\usepackage{amssymb}

\expandafter\let\csname equation*\endcsname\relax
\expandafter\let\csname endequation*\endcsname\relax
\usepackage{amsmath}
\usepackage{graphicx}% Include figure files
\usepackage{dcolumn}% Align table columns on decimal point
\usepackage{bm}% bold math
\usepackage{subcaption}
\usepackage{ragged2e}
\usepackage{tikz}
\usepackage{tikz-3dplot}
\usetikzlibrary{positioning}
\usetikzlibrary{calc,fadings,decorations.pathreplacing}
\usetikzlibrary{arrows,shapes,backgrounds}

\newcommand{\defn}{\stackrel{\textrm{\scriptsize def}}{=}}

\begin{document}

%\preprint{APS/123-QED}

%\submitto{\NJP}

\title[Sub-wavelength spin excitations in ultracold gases]{Sub-wavelength spin excitations in ultracold gases created by stimulated Raman transitions}

\author{Yigal Ilin$^1$, Shai Tsesses$^1$, Guy Bartal$^1$, Yoav Sagi$^{2*}$}
\address{$^1$ Andrew and Erna Viterbi Department of Electrical Engineering, Technion – Israel Institute of Technology, 3200003, Haifa, Israel}
\address{$^2$ Physics Department and Solid State Institute, Technion – Israel Institute of Technology, 3200003, Haifa, Israel}

\ead{$^{*}$yoavsagi@technion.ac.il}

%These lines are used for APS submission
%\author{Yigal Ilin}
%\author{Shai Tsesses}
%\author{Guy Bartal}
%\affiliation{Andrew and Erna Viterbi Department of Electrical Engineering, Technion – Israel Institute of Technology, 3200003, Haifa, Israel}
%\author{Yoav Sagi}
%\affiliation{Physics Department and Solid State Institute, Technion – Israel Institute of Technology, 3200003, Haifa, Israel}


\begin{abstract}
Raman transitions are used in quantum simulations with ultracold atoms for cooling, spectroscopy and creation of artificial gauge fields. Spatial shaping of the Raman fields allows local control of the effective Rabi frequency, which can be mapped to the atomic spin. Evanescent Raman fields are of special interest as they can provide a new degree of control emanating from their rapidly decaying profile and for their ability to generate features below the diffraction limit. This opens the door to the formation of sub-wavelength spin textures. In this work, we present a theoretical and numerical study of Raman Rabi frequency in the presence of evanescent driving fields. We show how spin textures can be created by spatially varying driving fields and demonstrate a skyrmionium lattice - a periodic array of topological spin excitations, each of which is composed of two skyrmions with opposite topological charges. Our results pave the way to quantum simulation of spin excitation dynamics in magnetic materials, especially of itinerant spin models. 
\end{abstract}
\noindent{\it Keywords}: Raman transitions, Ultracold atoms, Near field, Skyrmion, Spin texture, Skyrmionium, Quantum simulation, Sub-wavelength, Topological excitations
%\maketitle
%\tableofcontents
\subsection*{\label{sec:level1}\protect}
\section{Introduction}
The experimental science of ultracold atoms has proven to be instrumental in understanding complex many-body phenomena \cite{RevModPhys.80.885}. Furthermore, the near-perfect isolation from the environment and unrivaled controllability of neutral atoms has made them an ideal platform for quantum computation \cite{Bloch2012} and simulation \cite{Georgescu2014}. In this regard, it is possible to engineer the potential landscape of neutral atoms using far-off-resonance light \cite{Grimm200095},  and to simulate effects like the Lorentz force \cite{Lin2009} and spin-orbit coupling \cite{Lin2011} by coupling different spin states of the atom using Raman beams. Since the above-mentioned techniques utilize propagating optical fields,  spatial variations are naturally limited to $\sim\lambda/2$, where $\lambda$ is the wavelength of the light. Therefore, atoms cannot be brought closer than $\sim\lambda/2$ in a controlled manner, setting a constraint on the typical interaction energy and putting many interesting collective phases beyond experimental reach in the current regime of lowest achievable temperatures. 

Nanophotonics provides a promising route for generating spatial variations smaller than the diffraction limit, by applying in-plane wave-vectors larger than those available in free-space far-field propagation. In this approach, the fields decay exponentially relative to the nanophotonic device, requiring that the atoms be brought very close to the surface \cite{Gonzalez-Tudela2015,Chang2018}. Such fields appear, for example, in guided modes propagating at the boundary of thin metallic films \cite{Economou1969}, wave-guides \cite{Midwinter}, or in total internal reflection at an interface between two dielectric materials \cite{Burghardt1984}.

Integrating ultracold atoms in the near-field of nanophotonic devices not only gives access to sub-wavelength spatial variations of the fields, but also to the unique behavior of evanescent electromagnetic waves. In the near-field regime, radiation exhibits intriguing topological properties, e.g. spin-momentum locking or transverse spin \cite{Petersen2014,VanMechelen2016, Kalhor2016}, as well as topologically nontrivial patterns, the likes of optical skyrmions \cite{Tsesses2018a}. The latter are topologically stable field configurations \cite{Skyrme1962}, which have aroused great interest in the field of magnetism, as a unique and possibly applicable form of spin arrangement in solids \cite{Mhlbauer915, Yu2010, Nagaosa2013}. Though previous works studied Raman coupling in the near-field regime in the context of spectroscopy and imaging \cite{Hartschuh2003, Kawata2009}, the use of Raman transitions for coherent manipulation of the atomic spin degree of freedom in the near-field remains so far unexplored.

Here, we explore the use of Raman transitions in near-field evanescent electromagnetic waves for sub-wavelength spin structure generation in an ultracold atom gas. We derive a generalized formalism for the Raman Rabi frequency in the presence of an arbitrary evanescent driving field and discover a new method to induce spin excitations in the quantum gas. As a proof-of-concept, we consider the excitation of a skyrmionium lattice \cite{Zhang2016} - a topological excitation that carries a trivial total topological charge ($S=0$), but can be viewed as a combination of two concentric skyrmions with opposite topological charges of $S=\pm1$. We further discuss its experimental feasibility, and how such spin textures could enable quantum simulation of magnetic materials and the study of their dynamics.

\begin{figure}[ht]
	\centering
	\includegraphics[width=0.6\linewidth]{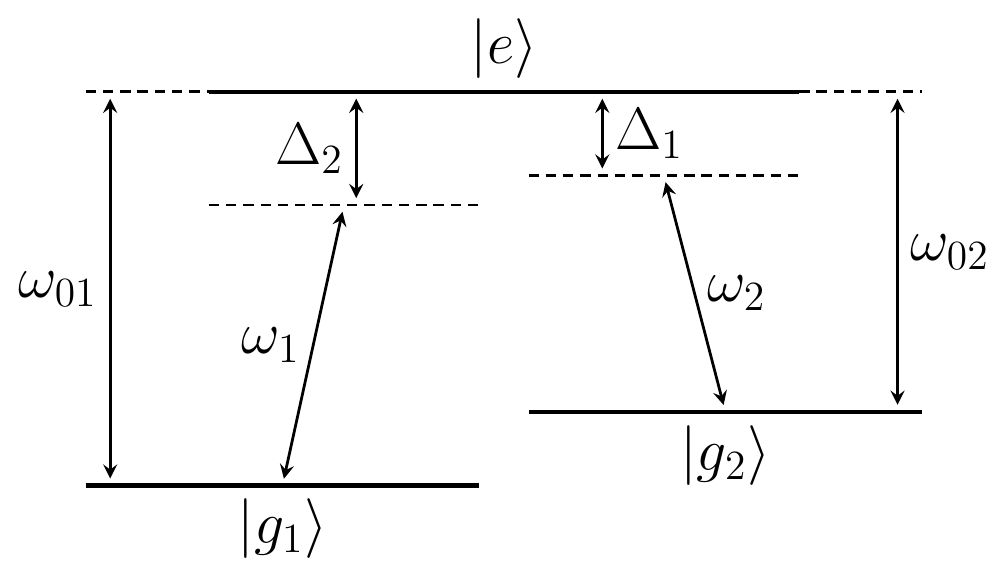}
    %\captionsetup{singlelinecheck=off,justification=raggedright} 
    \caption{The atomic energy levels diagram. Each atom can be viewed as a three-level system in a $\Lambda$ configuration. The ground states $|g_{1}\rangle$ and $|g_{2}\rangle$ are coupled to the excited state $|e\rangle$ by evanescent electric fields $\vec{E}_{1}$ and $\vec{E}_{2}$, respectively.}
	\label{Fig1}
\end{figure}

\section{Atom-field interaction model}
We consider an atom near a surface, interacting with an evanescent electric field. The field may be created in many ways, e.g. total internal reflection, and can be written as $\vec{E}(\vec{r})=\vec{E}_{1}(\vec{r})+\vec{E}_{2}(\vec{r})=\vec{\varepsilon}_{1}(x,y)e^{-|k_{1,z}|z}e^{j\omega_{1}t}+\vec{\varepsilon}_{2}(x,y)e^{-|k_{2,z}|z}e^{j\omega_{2}t}$, where $k_{\alpha,z}$ $(\alpha=1,2)$ is a purely imaginary wave vector component and $\vec{\varepsilon}_{\alpha}$ is a three dimensional vector field that depends on $(x,y)$ coordinates. The origin resides on the surface and the $\hat{z}$ axis is perpendicular to the surface. The fields $\vec{E}_{\alpha}$ have components both perpendicular (out-of-plane) and parallel (in-plane) to the surface. Each of the atoms is modeled as a three-level system in a $\Lambda$ configuration, as depicted in Fig. \ref{Fig1} \cite{Wu1996}. Collective effects due to close proximity of the atoms may be non-negligible, but we defer their inclusion to future works.

Under a dipole approximation \cite{alma990021232340203971}, the interaction Hamiltonian is given in the rotating frame by
%\begin{equation}\label{Eq_H_AF_1}
%\begin{split}
%    H_{AF} = -\frac{1}{2}\langle g_1|\vec{d}|e\rangle %\hat{\sigma}_1\cdot\vec{\tilde{E_1}}^{*}(\vec{r})\\- %\frac{1}{2}\langle g_2|\vec{d}|e\rangle %\hat{\sigma}_2\cdot\vec{\tilde{E_2}}^{*}(\vec{r})+h.c. \ \ ,
%\end{split}
%\end{equation}
\begin{equation}\label{Eq_H_AF_1}
    H_{AF} = -\frac{1}{2}\langle g_1|\vec{d}|e\rangle \hat{\sigma}_1\cdot\vec{\tilde{E_1}}^{*}(\vec{r})- \frac{1}{2}\langle g_2|\vec{d}|e\rangle \hat{\sigma}_2\cdot\vec{\tilde{E_2}}^{*}(\vec{r})+h.c. \ \ ,
\end{equation}
Where $\vec{\tilde{E}}_{\alpha}(\vec{r})$ is the electric field phasor, such that $\vec{E}_{\alpha}(\vec{r})=Re\left\{\vec{\tilde{E}}_{\alpha}(\vec{r})e^{j\omega_{\alpha}t}\right\}$ and  $\hat{\sigma}_\alpha=|g_{\alpha}\rangle\langle e|$ is the transition operator. Moving to spherical unit-vectors \cite{alma9926393111403971}: $ \hat{u}_{\pm1} = (\hat{x}\pm i\hat{y})/\sqrt{2}$ and $\hat{u}_{0}=\hat{z}$, each term in Eq.(\ref{Eq_H_AF_1}) becomes:
%\begin{equation}\label{eq_coupling_terms}
%\begin{split}
%    &\langle g_\alpha|\vec{d}|e\rangle\cdot\vec{\tilde{E}}_{\alph%a}(\vec{r}) =\\&\frac{1}{\sqrt{2}}\langle %g_\alpha|\vec{d}\cdot\hat{u}_{+1}|e\rangle\Big[\tilde{E}_{\alpha,%x}(\vec{r})-i\tilde{E}_{\alpha,y}(\vec{r})\Big]\\&+\frac{1}{\sqrt%{2}}\langle g_\alpha|\vec{d}\cdot\hat{u}_{-1}|e\rangle\Big[\tilde%{E}_{\alpha,x}(\vec{r})+i\tilde{E}_{\alpha,y}(\vec{r})\Big]\\&+\l%angle g_\alpha|\vec{d}\cdot\hat{u}_{0}|e\rangle\tilde{E}_{\alpha,%z}(\vec{r}) \ \ .
%\end{split}
%\end{equation}
\begin{equation}\label{eq_coupling_terms}
\begin{split}
    &\langle g_\alpha|\vec{d}|e\rangle\cdot\vec{\tilde{E}}_{\alpha}(\vec{r}) =\frac{1}{\sqrt{2}}\langle g_\alpha|\vec{d}\cdot\hat{u}_{+1}|e\rangle\Big[\tilde{E}_{\alpha,x}(\vec{r})-i\tilde{E}_{\alpha,y}(\vec{r})\Big]\\&+\frac{1}{\sqrt{2}}\langle g_\alpha|\vec{d}\cdot\hat{u}_{-1}|e\rangle\Big[\tilde{E}_{\alpha,x}(\vec{r})+i\tilde{E}_{\alpha,y}(\vec{r})\Big]+\langle g_\alpha|\vec{d}\cdot\hat{u}_{0}|e\rangle\tilde{E}_{\alpha,z}(\vec{r}) \ \ .
\end{split}
\end{equation}
It is evident from this expression that the coupling strength between each of the ground states and the excited state depends on the interference pattern created by the driving fields $\vec{E}_{\alpha}$. Each of the matrix elements in Eq.(\ref{eq_coupling_terms}) can be calculated using the Wigner 3j \cite{Aquilanti2007} and 6j \cite{Aquilanti2012} symbols \cite{alma9926393111403971, SUPPLEMENTARY}. The definition of the Rabi frequency for each transition $|g_{\alpha}\rangle \leftrightarrow |e\rangle$ reads
\begin{equation}\label{eq_generalized_rabi_freq}
    \Omega_{\alpha}(\vec{r}) = -\frac{1}{\hbar}\langle g_\alpha|\vec{d}|e\rangle\cdot\vec{\tilde{E}}_{\alpha}^{*}(\vec{r}) \ \ , 
\end{equation}
and is spatially dependent according to the driving field’s interference pattern. Using the definition of Eq.(\ref{eq_generalized_rabi_freq}), the atom-field interaction of Eq.(\ref{Eq_H_AF_1}) is written as:
\begin{equation}\label{eq_H_AF_2}
\begin{split}
    \tilde{H}_{AF} = \frac{\hbar \Omega_{1}(\vec{r})}{2}\hat{\sigma}_{1}+\frac{\hbar \Omega_{2}(\vec{r})}{2}\hat{\sigma}_{2}+h.c. \ \ .
\end{split}  
\end{equation}
Under the Born–Oppenheimer approximation \cite{Combes1981}, we write the atomic wave-function as a series of tensor products of internal and external states: $|\psi\rangle = |\psi_{g1}\rangle|g_{1}\rangle+|\psi_{g2}\rangle|g_{2}\rangle+|\psi_{e}\rangle|e\rangle$. We substitute this expression into Schrodinger’s equation and apply adiabatic elimination \cite{Brion2007}, thus arriving at the following equations of motion:
%\begin{eqnarray}
%    i\hbar\partial_{t}|\psi_{g1}\rangle &=& %\Big[\frac{\vec{p}\,^2}{2m}+\hbar\Delta_{1}+\frac{\hbar}{4\Delta}%\Big|\Omega_{1}(\vec{r})\Big|^{2}\Big]|\psi_{g1}\rangle \notag %\\&&+\hbar \Omega_{R}(\vec{r})|\psi_{g2}\rangle \notag \\
%    i\hbar\partial_{t}|\psi_{g2}\rangle &=& %\Big[\frac{\vec{p}\,^2}{2m}+\hbar\Delta_{2}+\frac{\hbar}{4\Delta}%\Big|\Omega_{2}(\vec{r})\Big|^{2}\Big]|\psi_{g2}\rangle \notag %\\&&+\hbar \Omega_{R}^{*}(\vec{r})|\psi_{g1}\rangle \ \ %,\label{eq_of_motion}
%\end{eqnarray}
\begin{eqnarray}\label{eq_of_motion}
    i\hbar\partial_{t}|\psi_{g1}\rangle &=& \Big[\frac{\vec{p}\,^2}{2m}+\hbar\Delta_{1}+\frac{\hbar}{4\Delta}\Big|\Omega_{1}(\vec{r})\Big|^{2}\Big]|\psi_{g1}\rangle \notag +\hbar \Omega_{R}(\vec{r})|\psi_{g2}\rangle \notag \\
    i\hbar\partial_{t}|\psi_{g2}\rangle &=& \Big[\frac{\vec{p}\,^2}{2m}+\hbar\Delta_{2}+\frac{\hbar}{4\Delta}\Big|\Omega_{2}(\vec{r})\Big|^{2}\Big]|\psi_{g2}\rangle +\hbar \Omega_{R}^{*}(\vec{r})|\psi_{g1}\rangle \ \ ,
\end{eqnarray}
where $\Omega_{R}$ is the generalized Raman Rabi frequency defined by
\begin{equation}\label{eq_generalized_Raman_Rabi_freq}
    \Omega_{R}(\vec{r}) = \frac{\Omega_{1}(\vec{r})\Omega_{2}^{*}(\vec{r})}{4\Delta} \ \ .
\end{equation}
$\Delta\defn \frac{\Delta_{1}+\Delta_{2}}{2}$ is the average single photon detuning, and we assume $\Delta \gg \{|\Delta_{1}-\Delta_{2}|,\,\Gamma\}$, with $\Gamma$ being the natural linewidth of the excited state. The r.h.s of Eqs.(\ref{eq_of_motion}) is divided into two parts: the free evolution of each ground state, and its coupling to the other ground state due to the Raman fields. The free evolution includes three terms that account for the kinetic energy, the bare ground state energy, and the light-shift due to the Raman field. The Raman coupling term is shaped by the interference pattern of the two Raman fields, which in turn translates into the ground state populations at each point.  

%\subsection*{}
\section{Selection rules}
Next, we seek the conditions under which the Raman coupling does not vanish. The coupling is proportional to $\Omega_{R}(\vec{r})\sim\langle g_{1}|\vec{d}\cdot\hat{u}_{q_{1}}|e\rangle\langle e|\left(\vec{d}\cdot\hat{u}_{q_{2}}\right)^{\dagger}|g_{2}\rangle$, where $q_{\alpha}=0,\pm1$ is the polarization vector for the corresponding driving field components in Eq.(\ref{eq_coupling_terms}). Let us assume that $|g_{1}\rangle$ and $|g_{2}\rangle$ are Zeeman states with $M_{F,1} = M_{F,2} - 1$. The Wigner 3j symbol implies momentum conservation which requires $q_{2}=q_{1}-1$. Combining these expressions each term of the form $\langle g_{1}|\vec{d}\cdot\hat{u}_{q_{1}}|e\rangle\langle e|\left(\vec{d}\cdot\hat{u}_{q_{2}}\right)^{\dagger}|g_{2}\rangle$ can be expanded as:
\begin{equation}\label{eq_matrix_elements}
\begin{split}
    &\langle g_{1}|\vec{d}\cdot\hat{u}_{q_{1}}|e\rangle\langle e|\left(\vec{d}\cdot\hat{u}_{q_{2}}\right)^{\dagger}|g_{2}\rangle = q_{e}^{2}\Big|\langle n_{s}^{*}L_{1}||r||n_{p}^{*}L_{e}\rangle\Big|^{2}\times(-1)^{1+2[L_{1}+S_{e}+J_{e}+J_{1}+I-M_{F,1}]}\\&\times[(2J_{e}+1)(2J_{1}+1)(2F_{e}+1)(2F_{1}+1)]
    \times\begin{Bmatrix}
    L_{1} & J_{1} & S_{e}\\
    J_{e} & L_{e} & 1
    \end{Bmatrix}^{2}
    \begin{Bmatrix}
    J_{1} & F_{1} & I\\
    F_{e} & J_{e} & 1
    \end{Bmatrix}^{2}\\&
    \times\begin{pmatrix}
    F_{e} & 1 & F_{1}\\
    M_{F,e} & q_{1} & M_{F,1}
    \end{pmatrix}
    \begin{pmatrix}
    F_{e} & 1 & F_{1}\\
    M_{F,e} & q_{1}-1 & M_{F,1}+1
    \end{pmatrix} \ \ ,
\end{split}
\end{equation}
where $N_{j=1,e}$ correspond to the quantum numbers of the states $|g_{1}\rangle$ and $|e\rangle$ respectively. Since $q_{1}=0,\pm 1$ we obtain that $\langle g_{1}|\vec{d}\cdot\hat{u}_{q_{1}}|e\rangle\langle e|\left(\vec{d}\cdot\hat{u}_{q_{2}}\right)^{\dagger}|g_{2}\rangle\neq 0$ only for $(q_{1},q_{2})\in \{(+1,0),(0,-1)\}$ \cite{SUPPLEMENTARY}. The Raman Rabi frequency then becomes
%\begin{equation}\label{eq_raman_rabi_2}
%\begin{split}
%    \Omega_{R}(\vec{r}) = %M_{+1,0}\Big[\tilde{E}_{1,x}(\vec{r})-i\tilde{E}_{1,y}(\vec{r})\B%ig]\tilde{E}_{2,z}^{*}(\vec{r})&\\+M_{0,-1}\Big[\tilde{E}_{2,x}^{%*}(\vec{r})-i\tilde{E}_{2,y}^{*}(\vec{r})\Big]\tilde{E}_{1,z}(\ve%c{r})
%\end{split}
%\end{equation}
\begin{equation}\label{eq_raman_rabi_2}
    \Omega_{R}(\vec{r}) = M_{+1,0}\Big[\tilde{E}_{1,x}(\vec{r})-i\tilde{E}_{1,y}(\vec{r})\Big]\tilde{E}_{2,z}^{*}(\vec{r})+M_{0,-1}\Big[\tilde{E}_{2,x}^{*}(\vec{r})-i\tilde{E}_{2,y}^{*}(\vec{r})\Big]\tilde{E}_{1,z}(\vec{r})
\end{equation}
Where $M_{q_1,q_2} = \frac{1}{4\sqrt{2}\Delta\hbar^{2}}\langle g_{1}|\vec{d}\cdot\hat{u}_{q_{1}}|e\rangle\langle e|\left(\vec{d}\cdot\hat{u}_{q_{2}}\right)^{\dagger}|g_{2}\rangle$. Eq.(\ref{eq_raman_rabi_2}) shows that Raman transitions are generated by a combination of the $(x,y)$ components of the $\vec{E}_{1}$ ($\vec{E}_{2}$) field and the $z$ component of the $\vec{E}_{2}$ ($\vec{E}_{1}$) field.

\begin{figure}[t]
    \includegraphics[width=\linewidth]{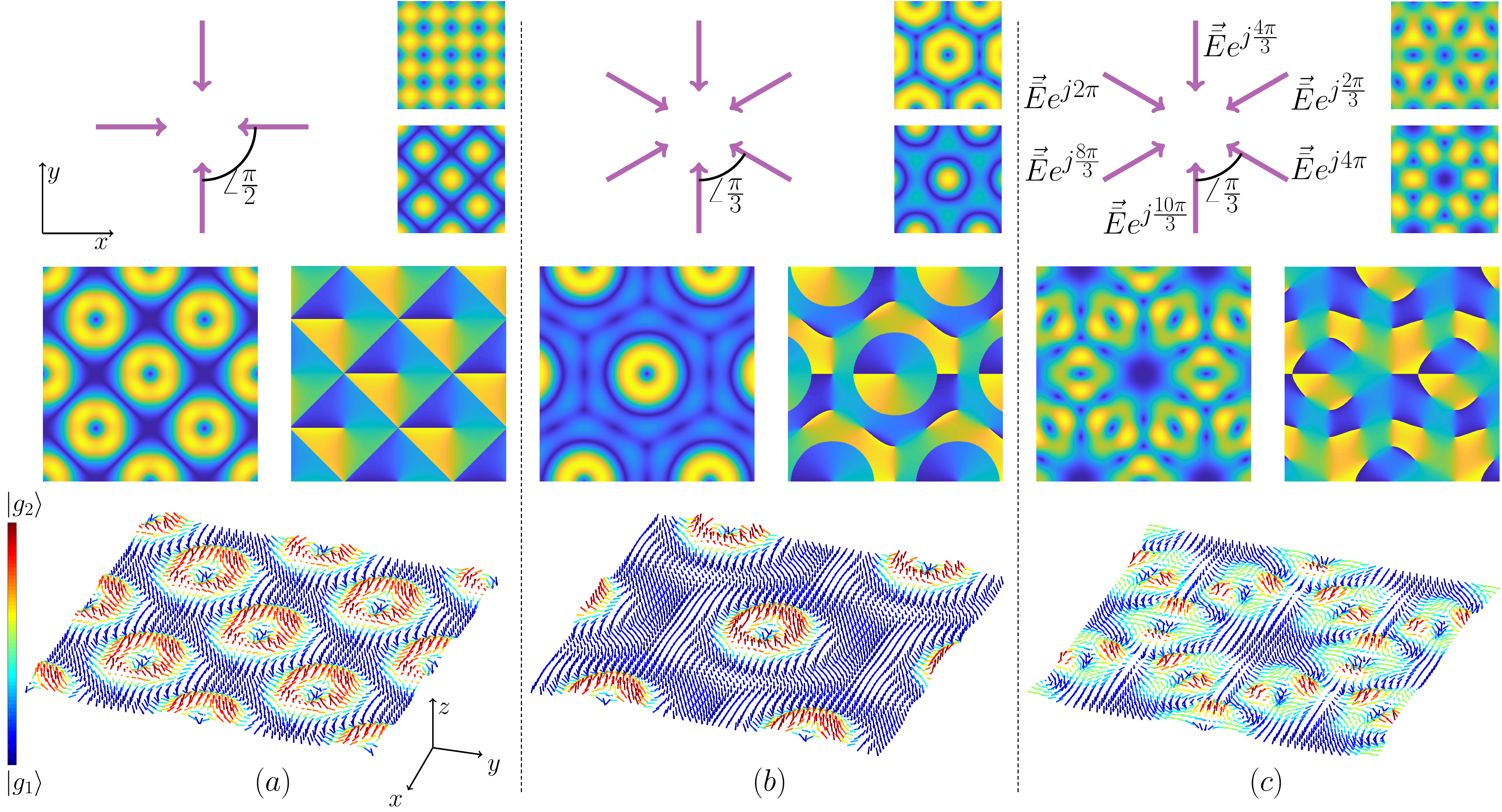}
    %\captionsetup{singlelinecheck=off,justification=raggedright}
    \caption{Spin textures generated by near-field Raman transitions. \textbf{Top row:} Field configuration and interference patterns. Arrows (left) represent the in-plane propagation direction, each arrow corresponds to both $\vec{E}_{1}$ and $\vec{E}_{2}$. Amplitude of the in-plane (upper right) and the out-of-plane (lower right) field components are shown next to the arrows. Bright (Dark) yellow (blue) denotes the maximal (minimal) value. Each configuration consists of two groups of counter-propagating beams, organized as (a) four beams in a square formation; (b) six beams in a hexagonal formation; and (c) six beams the same as in (b), but with different relative phases. \textbf{Middle row:} Spatially dependent Raman Rabi Frequency.  Amplitude (left) and phase (right) are shown for each configuration in the top row, resulting in a square (a), hexagonal (b) or Kagome (c) lattice.  Bright (Dark) yellow (blue) denoting the maximal (minimal) value. \textbf{Lower row:} Spin excitation in an atom cloud after Raman fields interact with the atoms for a duration $T=\pi/\underset{\vec{r}}{max}\Big(\Big|\Omega_{R}(\vec{r})\Big|\Big)$ ("$\pi$-pulse"). Spatial changes in coupling frequency affect the spinor state according to the mapping in Eq. (\ref{eq_raman_bloch_map}), such that the resulting spin texture follows the Raman Rabi frequency amplitude closely. Colorbar shows the state of the spinor as a function of position within the cold atom cloud. For these simulations we used typical numbers of ${}^{40}$K atoms. The Raman fields were taken to be red-detuned, $\Delta\approx-10THz$, from the $\text{D}_2$ transition (${}^{2}S_{1/2}\rightarrow{}^{2}P_{3/2}$). Spinor states are defined as $|g_{1}\rangle=|M_{F}=-9/2\rangle$ and $|g_{2}\rangle=|M_{F}=-7/2\rangle$ in the $F=9/2$ hyperfine manifold. The width of the insets is $1.8\lambda_{0}$, where $\lambda_{0}\approx 766.7$nm is the $\text{D}_2$ transition wavelengh.}
    \label{Fig2}
\end{figure}

%\subsection*{}
\section{Sub-wavelength spin excitations}
A major outcome of the position dependent Raman coupling in the near-field, given by Eq.(\ref{eq_raman_rabi_2}), is the possible generation of complex spin excitations. We assume that the atoms are held in a two-dimensional trap at some distance $z_0$ from the surface, and that their initial state is $|\varphi\rangle=|g_{1}\rangle$. We also assume that the Raman fields are pulsed for a duration $T$ much shorter than the dynamical timescales of the free evolution. We can therefore focus only on the coupling term in Eq.(\ref{eq_of_motion}), and leave the analysis of itinerant spins to future works. Each of the atoms is an effective two-level system that can be mapped to a spinor: 
\begin{equation}\label{eq_spin_state}
\begin{split}
    |\varphi\rangle = e^{-i\phi}\cos\left(\frac{\theta}{2}\right)|\uparrow\rangle+\sin\left(\frac{\theta}{2}\right)|\downarrow\rangle \ \ ,
\end{split}
\end{equation}
with $|\downarrow\rangle\defn|g_{1}\rangle$ and $|\uparrow\rangle\defn|g_{2}\rangle$ and the angles given by
\begin{equation}\label{eq_raman_bloch_map}
    \theta = 2\arccos{\left(\frac{T}{\pi}\Big|\Omega_{R}(\vec{r})\Big|\right)} \quad \quad \quad
    \phi = \measuredangle\Omega_{R}(\vec{r}) \ \ .
\end{equation}
%\begin{equation}\label{eq_raman_bloch_map}
%\begin{split}
%    \theta &= 2\arccos{\left(\frac{T}{\pi}\Big|\Omega_{R}%(\vec{r})\Big|\right)}\\
%    \phi &= \measuredangle\Omega_{R}(\vec{r}) \ \ .
%\end{split}
%\end{equation}
We take the duration $T=\pi/\underset{\vec{r}}{\text{max}}\Big(\Big|\Omega_{R}(\vec{r})\Big|\Big)$ such that spinors at the points with maximal Raman coupling are brought to the south pole of the Bloch sphere while atoms at points with vanishing $\Omega_{R}(\vec{r})$ will stay at the north pole. For this choice, Eqs. (\ref{eq_spin_state}-\ref{eq_raman_bloch_map}) map the spatially-varying Raman Rabi frequency onto a Bloch sphere, similarly to a spin-$1/2$ particle. In reality, the atoms are not necessarily spin-$1/2$ particles, but at high magnetic field the second-order Zeeman splitting induced between the other states is sufficiently far-detuned from the Raman resonance to neglect them. Thus, we can treat a multi-level atom as an effective spin-$1/2$ particle, as long as we isolate the Raman coupling to only two states.

We are now ready to calculate the resulting sub-wavelength spin excitation for several field configurations. In these calculations we assume ${}^{40}$K atoms are suspended in a two-dimensional trap at a distance of $z_{0}\sim100$nm from the surface. The surface is illuminated by $2n$ beams undergoing a total internal reflection. These beams are the source for the evanescent field that drives the Raman transitions at $z>0$. The phasor of a transverse-magnetic (TM) driving field $\vec{E}_{\alpha}$ is given by:
%\begin{equation}\label{eq_field_config}
%\begin{split}
%    &\vec{\tilde{E}}_{\alpha}(\vec{r}) = %E_{0,\alpha}e^{-|k_{\alpha,z}|z}\sum_{i=1}^{n}
%    \begin{pmatrix}
%    \frac{|k_{\alpha,z}|}{k_{\alpha,\|}}A_{\alpha,i} &\\
%    \frac{|k_{\alpha,z}|}{k_{\alpha,\|}}B_{\alpha,i} &\\
%    1 &
%    \end{pmatrix}\\
%    &\times e^{j\nu_{\alpha,i}}e^{-jk_{\alpha,\|}\left(A_{\alpha,%i}x+B_{\alpha,i}y\right)} \ \ ,
%\end{split}
%\end{equation}
\begin{equation}\label{eq_field_config}
    \vec{\tilde{E}}_{\alpha}(\vec{r}) = E_{0,\alpha}e^{-|k_{\alpha,z}|z}\sum_{i=1}^{n}
    \begin{pmatrix}
    \frac{|k_{\alpha,z}|}{k_{\alpha,\|}}A_{\alpha,i} &\\
    \frac{|k_{\alpha,z}|}{k_{\alpha,\|}}B_{\alpha,i} &\\
    1 &
    \end{pmatrix}\times e^{j\delta_{\alpha,i}}e^{-jk_{\alpha,\|}\left(A_{\alpha,i}x+B_{\alpha,i}y\right)} \ \ ,
\end{equation}
where $E_{0,\alpha}$ is the driving field’s amplitude (assumed to be real), $k_{\alpha,\|}=\sqrt{k_{\alpha,x}^{2}+k_{\alpha,y}^{2}}$ is the in-plane wave-vector, $k_{\alpha,0}$ is the free space wave-vector satisfying ${k_{\alpha,0}^{2}=k_{\alpha,\|}^{2}+k_{\alpha,z}^{2}}$, $\delta_{\alpha,i}$ is a constant initial phase of the beam $i$ and $A_{\alpha,i}$ and $B_{\alpha,i}$ are unit-vector components that define the in-plane propagation direction of the beam $i$.

To demonstrate spin texture shaping, we focus on three different driving field configurations, as shown in Fig. \ref{Fig2}. As can be seen, the symmetry of the spin excitation lattice is directly related to the symmetry of the driving fields. The length scale on which the spins are rotated in the Bloch sphere is defined by the in-plane wave-vectors $k_{\alpha,\|}$ \cite{Tsesses2019}, controlled by the incidence angle and can be sub-wavelength with respect to the driving field’s free-space wavelength $\lambda_{\alpha,0}$.

%\subsection*{}
\section{Skyrmionium and antiskyrmionium lattices}
\begin{figure}[ht]
	\centering
	\includegraphics[width=0.95\linewidth]{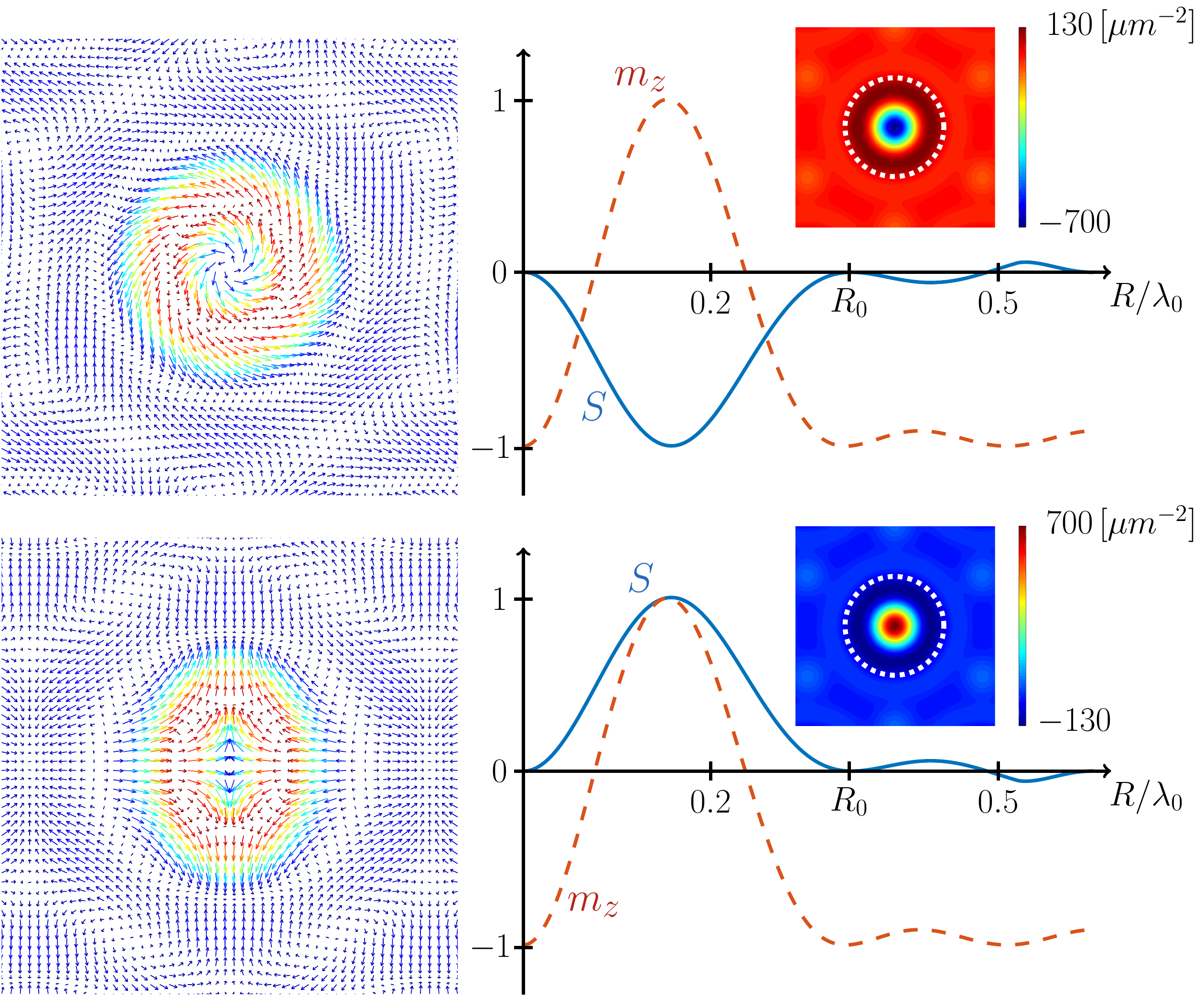}
	%\captionsetup{singlelinecheck=off,justification=raggedright}
    	\caption{Calculation of the topological charge as a function of the manifold $A$ from Eq. (\ref{eq_top_charge_integral}) and the $z$-component of the Bloch vector $\vec{m}$. \textbf{Left:} Unit cells in bottom and top rows correspond to the hexagonal beam formations presented in Fig. \ref{Fig2}(b) and Fig. \ref{Fig2}(c), where for the bottom row the relative phase between the beams is zero and for the top row the relative phase changed to be $\delta_{1(2),i}=+(-)\frac{i\pi}{3},\,i=1,..,6$ for $\vec{E}_{1}(\vec{E}_{2})$. \textbf{Right:} Numerical evaluation of $S$, given by Eq. \ref{eq_top_charge_integral}, as a function of $R$ that defines the integration domain and normalized by $\lambda_{0}$ (blue solid line). The value of $m_{z}$, the $z$ component of the Bloch vector defined by Eq. \ref{eq_raman_bloch_map}, is depicted as a dashed red line. For the top (bottom) row, as the topological charge reaches $-1$ $(+1)$, the Bloch vector completes a rotation of $-\pi$. The topological charge vanishes for the integration domain defined for $R_{0}=0.348$ as the Bloch vector rotates back $\pi$. The boundary of the domain defined by $R_{0}$ is highlighted in the insets as a dashed white line. (Insets) Topological charge densities over which integration is performed. The topological charge densities for the top and bottom rows are of opposite sign and the same magnitude.}
	\label{Fig3}
\end{figure}

We focus on Fig. \ref{Fig2}(b) and note that the spinor undergoes a continuous rotation from $|g_{1}\rangle$ to $|g_{2}\rangle$ and then back to $|g_{1}\rangle$ as we traverse over a straight line from the center of a unit cell to the boundary. To characterize the topology of the spin structure, we consider the topological charge density defined as $s\defn\vec{m}\cdot\left(\frac{\partial\vec{m}}{\partial x}\times\frac{\partial\vec{m}}{\partial y}\right)$ where $\vec{m}=(\cos\phi \sin\theta,\,\sin\phi \sin\theta,\,\cos\theta)$ is the Bloch vector. The topological charge within a unit cell can be described by a Pontryagin number defined as \cite{Nagaosa2013, Liu2016}:
\begin{equation}\label{eq_top_charge_integral}
\begin{split}
    S = \frac{1}{4\pi}\int_A s\,dA \ \ ,
\end{split}
\end{equation}
where $A$ is the two-dimensional manifold. To investigate how the Pontryagin number changes as we expand the domain of integration $A$ from the center of a unit cell to its boundary, we define the domain $A\defn \left\{(x,y)\,|\,x^{2}+y^{2}\leq R\right\}\bigcap B$ where $B$ is unit cell boundary and $R$ is some radius of choice. This definition essentially takes the intersection of ever increasing circles and the boundary of a unit cell \cite{SUPPLEMENTARY}. Fig. \ref{Fig3} shows the topological charge calculated for the spin lattices shown in panels (b) and (c) of Fig. \ref{Fig2}. We identify the topological behavior depicted in the top row of Fig. \ref{Fig3} as skyrmionium: a spatial spin excitation with a total topological charge of $S=0$ which is a combination of two concentric Bloch-type skyrmions with opposite topological charges of $S=\pm1$ \cite{Bogdanov1999}. The lower row of Fig. \ref{Fig3} corresponds to an antiskyrmionium composed of two concentric antiskyrmions with opposite topological charges. Such spin excitations were only recently observed in solid-state systems \cite{Finazzi2013,Komineas2015,Komineas2015a,Zhang2016,Thorsten2018} and were suggested as a platform for magnetic logic that is immune to parasitic Hall effects \cite{Kolesnikov2018}.

%\subsection*{}
\section{Discussion and outlook}
Our analysis shows that when atoms are driven by evanescent fields, a Raman process can be used to generate a sub-wavelength lattice of spin excitations, with the skyrmionium lattice presented above as a specific example. This opens the door to the study of spin excitation dynamics in magnetic materials using ultracold gases, though there are several challenges in this route. First, it is required to efficiently load the ultracold gas to a trap very close to a surface, which is at a much higher temperature than the atoms - a process usually achieved only with low-loss wave-guides \cite{Rausch2010}. Another challenge is to find a method to characterize the state of the gas with an underlying structure at a lengthscale smaller than the diffraction limit. One possible route is by using super-resolution imaging methods for atoms \cite{PhysRevX.9.021001}. Finally, if the average distance between the atoms is considerably shorter than the wavelength, collective effects such as sub- and super-radiance may occur as well. These effects can also be potentially useful in probing and characterizing the atomic spin structure \cite{rui2020subradiant}.

The use of ultracold atoms allows studying interesting models of magnetism which are hard to implement in solid systems. For example, by trapping the atoms in an optical lattice (which may also be sub-wavelength in the near-field regime), one can explore a model of itinerant spins and the interplay between effective magnetic interactions and diffusion. In addition, Feshbach resonances enable tuning the interaction strength between the two spin states. We believe that the most interesting regime will be around unitarity, where the spin diffusivity achieves its minimum \cite{Sommer2011}. 

Another interesting direction is to study the dynamics of a skyrmionium lattice in the repulsive branch of the Feshbach resonance. The Stoner model predicts itinerant fermions with repulsive interactions will develop ferromagnetic ordering \cite{Snoke2009}, but rapid decay from the excited repulsively-interacting state prevented the experimental observation of this phenomenon thus far \cite{PhysRevLett.108.240404}. The rapid decay was due to the formation of bound pairs of particles with opposite spins. However, in a skyrmionium, the spatial spin rotation happens at a low wave-vector, hence adjacent spins are pointing almost to the same direction and Pauli repulsion is expected to suppress recombination processes. Therefore, it is plausible that in the repulsive branch a spin excitation will actually be more stable than in a fully mixed spin state. 

It is worth noting that our results bear some resemblance to the study of optical flux lattices in ultracold atomic gases, where spin textures consisting of both zero \cite{Price2011} and non-zero \cite{Kasamatsu2004, Cooper2011, Cooper2011a} topological charges were envisioned. In these works, spin textures were obtained by either minimization of the Gross-Pitaevskii energy functional \cite{Kasamatsu2004, Price2011} or by considering the eigenstates of the two-level system \cite{Cooper2011, Cooper2011a}. Another possible route for spin texture generation in ultracold atoms is the onset of a Rashba Hamiltonian in the interaction of atoms with an optical lattice \cite{Dudarev2004}. The derivation performed in this paper, however, offers a different treatment to this problem, which may be implemented using extension of existing techniques.

\ack
%\subsubsection*{Acknowledgments}
We wish to thank Ido Kaminer for fruitful discussions. This research was supported by a NEVET grant from Russell Berrie Nanotechnology Institute (RBNI) at the Technion - Israel Institute of Technology. S.T. acknowledges support by the Adams Fellowship Program of the Israel Academy of Science and Humanities.

\section*{References}
\bibliography{raman_near_field.bib}
\end{document}